\documentclass[conference]{IEEEtran}

\usepackage[dvips]{graphicx}
\usepackage{multirow}
\usepackage{footmisc}

\ifCLASSINFOpdf
\else
\fi

\usepackage[tight,footnotesize]{subfigure}
\usepackage{footnote}
\makesavenoteenv{tabular}

\begin{document}

\title{Preliminary Analysis of ULPC Light Curves Using Fourier Decomposition Technique}

\author{
\IEEEauthorblockN{Chow-Choong Ngeow\IEEEauthorrefmark{3}, Scott Lucchini\IEEEauthorrefmark{1}, Shashi Kanbur\IEEEauthorrefmark{2}, Brittany Barrett\IEEEauthorrefmark{2} and Bin Lin\IEEEauthorrefmark{1}}
\IEEEauthorblockA{\IEEEauthorrefmark{3}Graduate Institute of Astronomy, National Central University, Jhongli 32001, Taiwan }
\IEEEauthorblockA{\IEEEauthorrefmark{1}Department of Physics and Astronomy, University of Rochester, Rochester, NY 14627, USA }
\IEEEauthorblockA{\IEEEauthorrefmark{2}Department of Physics, SUNY Oswego, Oswego, NY 13126, USA }
}

\maketitle

\begin{abstract}

Recent work on Ultra Long Period Cepheids (ULPCs) has suggested their usefulness as a distance indicator, but has not commented on their relationship as compared with other types of variable stars. In this work, we use Fourier analysis to quantify the structure of ULPC light curves and compare them to Classical Cepheids and Mira variables. Our preliminary results suggest that the low order Fourier parameters ($R_{k1}$ and $\phi_{k1}$) of ULPCs show a continuous trend defined by Classical Cepheids after the resonance around 10 days. However their Fourier parameters also overlapped with those from Miras, which make the classification of long period variable stars difficult based on the light curves information alone.

\end{abstract}


\IEEEpeerreviewmaketitle

\section{Introduction}
\label{sec:intro}
Cepheid variables are key component in astrophysical research, as they are an indispensable link in the extragalactic distance ladder. However, for the most part, current research focuses on Classical Cepheids with periods less than 60 days. There are good reasons for this: in terms of sufficient phase coverage, they take less time to observe, and they have a well defined period-luminosity (PL) relation. Also, Cepheids with periods greater than 100 days obey a different relation (\cite{bird}), which has not been nearly as well documented.  In this work, we follow the same conventions as \cite{bird}, defining these ``Ultra Long Period Cepheids'' (ULPCs) to be any fundamental mode Cepheids with a pulsation period greater than 80 days.
\par
Even though ULPCs have generally not been studied as intensely as shorter period Cepheids, there are many advantages to acquiring data on these ULPCs.  As the PL relation indicates, the longer the period of a star, the brighter it is.  This means that these variables can be observed at much greater distances than are currently possible.  This added brightness would allow us to use Cepheids to measure distances out to 100 Mpc and beyond \cite{bird}. In this work, our goal is to study the structural properties of ULPC light curves.  We are interested in comparing the light curves of ULPCs to shorter period Classical Cepheids as well as to longer period Mira-like variables (Miras).

\begin{figure*}[!t]
\centering
\includegraphics[height=\textwidth, angle=-90]{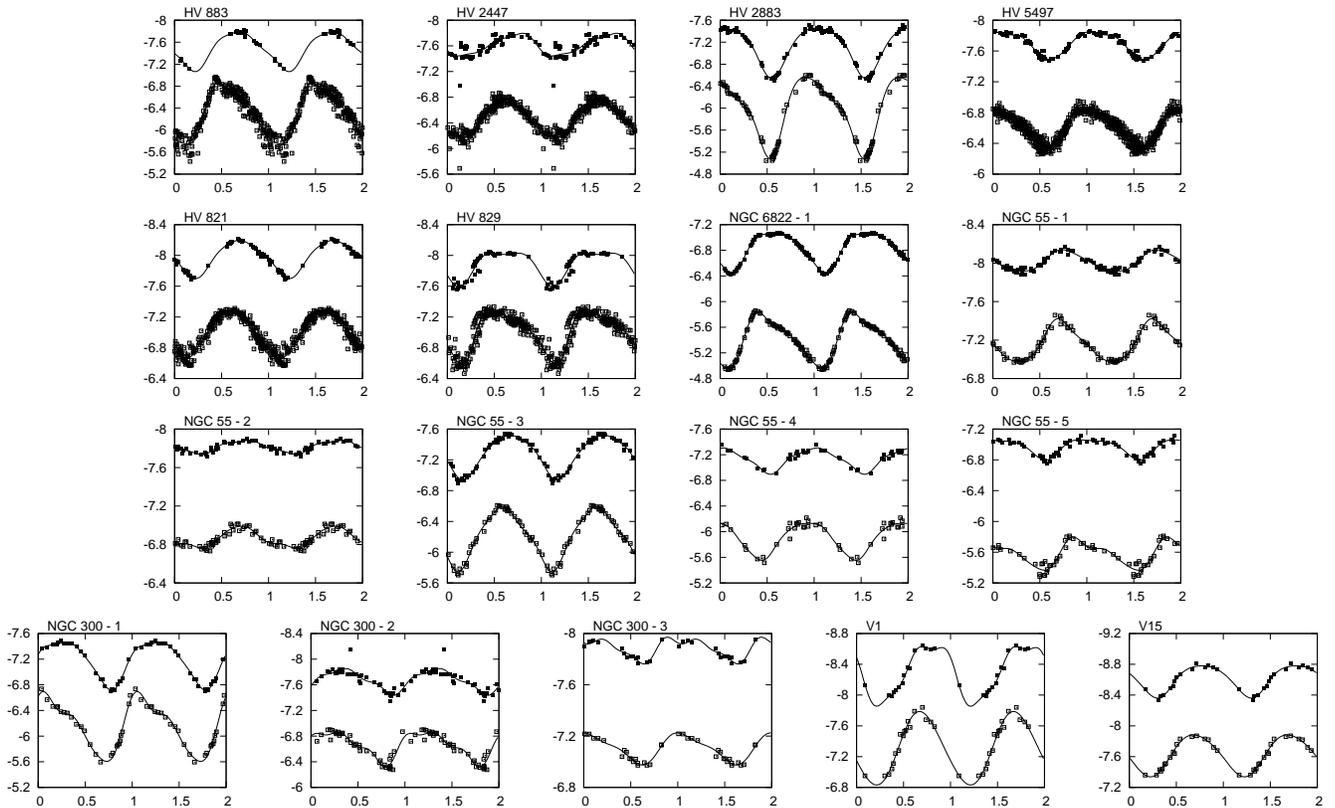}
\caption{ULPC light curves with their respective Fourier fits.  The absolute V (open squares) and I (filled squares) magnitudes are represented along the y-axis, and the phase of pulsation is along the x-axis.}
\label{fig:ulplcv}
\end{figure*}

\section[]{Data}
\label{sec:data}

\subsection{ULPCs}
\label{sec:ulpcs}
There is very limited data for ULPCs available right now, but we were able to compile data for 17 ULPCs in 6 different galaxies, including the Large and Small Magellanic Clouds (LMC/SMC), NGC 6822, NGC 55, NGC 300, and I Zw 18.
\par
For the six stars in the LMC and the SMC, we obtained the V-band data from \cite{bird} and the I-band data from \cite{moffett}.  We also acquired additional V- and I-band data through the McMaster Cepheid Photometry Archive\footnote{http://crocus.physics.mcmaster.ca/Cepheid/} for four ULPCs in the LMC.  \cite{freedman} had some additional data points for HV 883 and HV 2447.  Also, for HV 2883 and HV 5497, we acquired data from \cite{martin}, again through the McMaster Archive. \cite{bird} also had V- and I-band data for the five stars in NGC 55, the three stars in NGC 300, and the one star in NGC 6822.  Again, using the McMaster Archive, additional photometry was acquired for NGC 300-1 (as named by \cite{bird}) from \cite{freedman}. The photometric V- and I-band data for the two stars in I Zw 18 were obtained from \cite{fiorentino}.
\par
The periods for these stars were also compiled in \cite{bird}.  In Table \ref{tab:ulpcs} we have listed all the ULPCs in our data set together with their periods and host galaxies. In this Table, the names of the stars (Column 1) are consistent with those used by \cite{bird}.  Column 3 contains the distance modulus (DM) values for the stars' host galaxies listed in Column 2.  The magnitudes (Column 5) are the absolute, reddening corrected, mean V-band magnitudes as calculated by our Fourier analysis with orders as shown in Column 6.  All the data for these stars was obtained from \cite{bird} and references therein, however for the stars where more data was compiled, footnotes have been added. Figure \ref{fig:ulplcv} presents the ULPC light curves used in this work.

\begin{table}
\begin{center}
\begin{minipage}{\columnwidth}
\renewcommand{\thefootnote}{\thempfootnote}
\caption{Data on the ULPCs that we used.}
\centering
\begin{tabular}{@{}llrrrr@{}}
\hline
Name & Galaxy & DM & Period [days] & $M_V$ & N\\
\hline
HV 883\footnote{Additional data from \cite{freedman}.\label{fn:f}} & LMC & \multirow{4}{*}{18.50} & 133.6 & -6.30 & 4 \\
HV 2447\footref{fn:f} & LMC & & 118.7 & -6.47 & 4 \\
HV 2883\footnote{Additional data from \cite{martin}.\label{fn:m}} & LMC & & 109.2 & -5.98 & 3 \\
HV 5497\footref{fn:m} & LMC & & 98.6 & -6.62 & 4 \\
HV 821 & SMC & \multirow{2}{*}{18.93} & 127.5 & -6.99 & 4 \\
HV 829 & SMC & & 84.4 & -7.00 & 4 \\
NGC 6822 - 1 & NGC 6822 & 23.31 & 123.9 & -5.41 & 3 \\
NGC 55 - 1 & NGC 55 & \multirow{5}{*}{26.43} & 175.9 & -7.17 & 4 \\
NGC 55 - 2 & NGC 55 & & 152.1 & -6.87 & 3 \\
NGC 55 - 3 & NGC 55 & & 112.7 & -6.21 & 3 \\
NGC 55 - 4 & NGC 55 & & 97.7 & -5.89 & 4 \\
NGC 55 - 5 & NGC 55 & & 85.1 & -5.58 & 3 \\
NGC 300 - 1\footref{fn:f} & NGC 300 & \multirow{3}{*}{26.37} & 115.8 & -6.15 & 3 \\
NGC 300 - 2 & NGC 300 & & 89.1 & -6.66 & 3 \\
NGC 300 - 3 & NGC 300 & & 83.0 & -7.11 & 2 \\
V1 & I Zw 18 & \multirow{2}{*}{31.30} & 129.8 & -7.30 & 2 \\
V15 & I Zw 18 & & 125.0 & -7.62 & 2 \\
\hline
\label{tab:ulpcs}
\end{tabular}
\end{minipage}
\end{center}
Note: DM is the distance modulus of the host galaxy, $M_V$ is the absolute $V$-band averaged magnitude after reddening correction, and $N$ is the order of Fourier fit.
\end{table}

\subsection{OGLE-III Catalog}
\label{sec:ogle}
The third installment of the Optical Gravitational Lensing Experiment (OGLE-III) contains photometric data for many variable stars in the Large and Small Magellanic Clouds.  We were interested in the V- and I-band photometric light curves data for fundamental mode Cepheids and longer period Mira-like variables.  We also obtained the periods for all the stars we used from the catalog. From OGLE-III LMC Cepheid catalog (\cite{soslmc}), we chose 1,804 (out of 1,849) Cepheids based on what data was available for these stars.  We then obtained the V- and I-band photometric light curve data for these Cepheids. Similarly, of the 2,626 available fundamental mode Cepheids in the SMC (\cite{sossmc}), we selected 2,596. The OGLE-III catalog also contains photometric and period data for "long period variables" (LPVs).  These LPVs are classified into three different types: Miras, OSARGs, and SRVs (\cite{sos}).  We chose 1,407 of the listed 1,663 Mira variables and again used both their V- and I-band photometric data. 

\subsection{NGC 300 Cepheid Data}
\label{sec:ngc}
\cite{gieren} acquired V- and I-band photometric data for 64 Cepheids from NGC 300 with periods ranging from 11 to 115 days.  However, three of the Cepheids have periods greater than 80 days, so they are included as ULPCs, not as regular Cepheids of NGC 300.  The remaining 61 stars and their light curve data were added to our Classical Cepheid data set.

\section[]{Fourier Analysis}
\label{sec:fourier}
We characterized the structure of all the observed light curves in our data set using the technique of Fourier decomposition. That is, the following function was fit to the observed data for 
a given star:
\begin{equation}
\mathrm{mag}(t)=A_{0} + \sum_{k=1}^{N} A_{k} \sin(k\omega t + \phi_{k})
\label{eq:fourier}
\end{equation}
Here $\omega = {2\pi /P}$, where $P$ is the period in days, $A_k$ and $\phi_k$ represent the amplitude and phase-shift for $k^{\mathrm{th}}$-order respectively, and $N$ is the order of the fit.  To determine the order, we used several different techniques.  For the ULPCs, we looked at the Fourier fits of each star individually and visually determined which order gave the best representation.  The resulting orders are presented in last Column of Table \ref{tab:ulpcs}.  To determine the optimum orders for the Classical Cepheids and the Miras, we ran the Fourier analysis with several different orders and after inspecting some of the fits, concluded that the orders listed in Column 5 of Table \ref{tab:ccmiradata} gave the best representations of our data. This was a universal value that we used for all the stars of the given type in the given galaxy. In order to quantify the structure of the light curve, we used the Fourier parameters $R_{k1}$ and $\phi_{k1}$ defined as follow (\cite{simon}):
\begin{equation}
R_{k1} = \frac{A_{k}}{A_{1}}; \quad \phi_{k1} = \phi_{k} - k\phi_{1}
\label{eq:rk1def}
\end{equation}
where $k$ is set to be $2$ and $3$.

\begin{table}
\centering
\begin{minipage}{\columnwidth}
\renewcommand{\thefootnote}{\thempfootnote}
\caption{Information on the Classical Cepheids and Miras that we used in comparison with the ULPCs.}
\centering
\begin{tabular}{@{} l l r c r @{}}
\hline
Galaxy & Star Type & Stars & Band & Order \\
\hline
\multirow{4}{*}{LMC} & \multirow{2}{*}{Cepheids} & \multirow{2}{*}{1,804} & V & 5 \\
 & & & I & 8 \\
 & \multirow{2}{*}{Miras} & \multirow{2}{*}{1,407} & V & 3  \\
 & & & I & 3 \\
\multirow{2}{*}{SMC} & \multirow{2}{*}{Cepheids} & \multirow{2}{*}{2,598} & V & 4 \\
 & & & I & 8 \\
\multirow{2}{*}{NGC 300} & \multirow{2}{*}{Cepheids} & \multirow{2}{*}{61} & V & 3 \\
 & & & I & 3 \\
\hline
\label{tab:ccmiradata}
\end{tabular}
\end{minipage}
\end{table}

\subsection{Error on Fourier Parameters}
\label{sec:fouriererr}
In order to determine the accuracy of our results, we used techniques developed by \cite{petersen} to calculate the errors on $R_{k1}$ and $\phi_{k1}$ parameters as follows:
\begin{equation}
\sigma^2(R_{k1})=A_{1}^{-4}\epsilon^2(A_1^2+A_k^2)
\end{equation}
\begin{equation}
\sigma^2(\phi_{k1})=\epsilon^2(A_k^{-2}+k^2 A_1^{-2})
\end{equation}
where $A$ is the coefficient of the Fourier decomposition as shown in Equation (\ref{eq:fourier}) and $\epsilon$ is related to the sum of the squared residuals, $[vv]$:
\begin{equation}
\epsilon^2=\frac{2}{J}\frac{[vv]}{J-2N-1}, \quad \mathrm{and}\quad[vv]=\sum_{j=1}^Jv_jv_j
\end{equation}
In the above equation, $J$ is the number of data points, and $v_j$ is the $j^{th}$ residual.  Even though these are the approximations that \cite{petersen} derived, he states that they still provide valuable information on the accuracy of the Fourier parameters.

\begin{figure}
\centering
\includegraphics[height=\columnwidth, angle=-90]{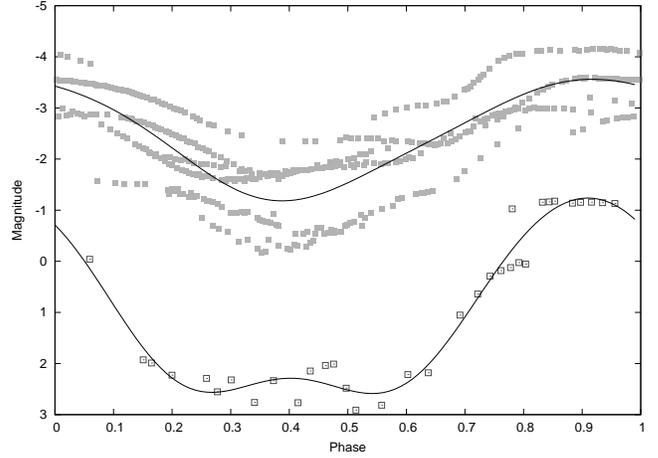}
\caption{The graph of the V (open squares) and I (solid squares) band light curves for OGLE-LMC-LPV-01408 as classified in the OGLE-III database.  Note that the I-band data has large dispersion. This is because of the variation in period and amplitude of the Miras' light curves. The Fourier fit (solid black lines) is an approximate ``average'' of the data points.}
\label{fig:miralcv}
\end{figure}

\subsection{Fourier Fit to the Miras Data}
\label{sec:miraerr}
Miras are known to have periods and amplitudes that are not constant like the Classical Cepheids.  This made compiling the light curves slightly more difficult, because the data points did not always form a single, well-defined curve when the measurement times were converted to phases of one period.  The I-band data was especially problematic.  We dealt with this by using a low enough order of the Fourier fit (as given in Table  \ref{tab:ccmiradata}) so as to essentially take an average of the light curve. This is demonstrated in Figure \ref{fig:miralcv}, where the solid squares are the I-band measurements.  The solid black lines are calculated the Fourier fits: the I-band fit is approximately an average of the spread of data points. In order to calculate the Fourier parameters stated above, we needed at least third order terms (as for some of the Miras, a higher order fit produced some artifacts in fitted light curves). Therefore, when we calculated the Fourier parameters and plotted the fitted light curves, we restricted to a third order fit. 

\begin{table*}
\centering
\renewcommand{\thefootnote}{\thempfootnote}
\caption{Fitted Fourier parameters for ULPCs.}
\centering
\begin{tabular}{@{}lccccc@{}}
\hline
Name & $\log(P)$ & $R_{21}$ & $\phi_{21}$ & $R_{31}$ & $\phi_{31}$ \\
\hline
\multicolumn{6}{c}{V-band Results} \\
\hline
HV 883		& 2.126 & $0.230\pm0.003$ & $3.464\pm0.064$ & $0.103\pm0.003$ & $18.737\pm0.293$ \\ 
HV 2447		& 2.074 & $0.066\pm0.005$ & $3.495\pm1.204$ & $0.027\pm0.005$ & $\ast 13.280\pm7.182$ \\
HV 2883		& 2.038 & $0.342\pm0.005$ & $3.876\pm0.053$ & $0.095\pm0.004$ & $14.692\pm0.500$ \\
HV 5497		& 1.994 & $0.242\pm0.005$ & $3.766\pm0.094$ & $0.034\pm0.004$ & $12.617\pm3.840$ \\
HV 821		& 2.106 & $0.070\pm0.005$ & $4.175\pm0.975$ & $0.046\pm0.005$ & $14.026\pm2.261$ \\ 
HV 829		& 1.926 & $0.391\pm0.007$ & $3.935\pm0.061$ & $0.125\pm0.006$ & $14.065\pm0.419$ \\ 
NGC6822-1	& 2.093 & $0.345\pm0.004$ & $3.477\pm0.040$ & $0.104\pm0.003$ & $18.843\pm0.329$ \\ 
NGC55-1		& 2.245 & $0.185\pm0.024$ & $2.534\pm0.783$ & $0.080\pm0.024$ & $17.162\pm3.891$ \\ 
NGC55-2		& 2.182 & $0.227\pm0.108$ & $2.243\pm2.401$ & $0.099\pm0.104$ & $\ast 13.699\pm11.396$ \\
NGC55-3		& 2.052 & $0.139\pm0.010$ & $4.069\pm0.543$ & $0.110\pm0.010$ & $15.427\pm0.889$ \\ 
NGC55-4		& 1.990 & $0.160\pm0.087$ & $4.732\pm3.636$ & $0.040\pm0.085$ & $\ast 14.177\pm52.827$ \\ 
NGC55-5		& 1.930 & $0.417\pm0.120$ & $3.567\pm0.993$ & $0.138\pm0.104$ & $\ast 18.434\pm6.275$ \\ 
NGC300-1	& 2.064 & $0.339\pm0.022$ & $3.561\pm0.248$ & $0.117\pm0.020$ & $18.318\pm1.611$ \\ 
NGC300-2	& 1.950 & $0.310\pm0.055$ & $3.870\pm0.727$ & $0.273\pm0.054$ & $14.172\pm1.132$ \\ 
NGC300-3	& 1.919 & $0.289\pm0.169$ & $3.491\pm2.494$ & $\cdots$ & $\cdots$ \\ 
V1		& 2.113 & $0.072\pm0.027$ & $\ast 2.938\pm5.207$& $\cdots$ & $\cdots$ \\ 
V15		& 2.097 & $0.102\pm0.032$ & $3.985\pm3.198$ & $\cdots$ & $\cdots$ \\
\hline
\multicolumn{6}{c}{I-band Results} \\
\hline
HV 883		& 2.126 & $0.230\pm0.003$ & $3.991\pm0.771$ & $0.100\pm0.019$ & $14.480\pm2.085$ \\
HV 2447		& 2.074 & $0.066\pm0.005$ & $6.239\pm2.215$ & $0.155\pm0.075$ & $12.815\pm3.684$ \\
HV 2883		& 2.038 & $0.342\pm0.005$ & $4.593\pm0.101$ & $0.055\pm0.006$ & $15.922\pm1.968$ \\
HV 5497		& 1.994 & $0.242\pm0.005$ & $4.421\pm0.522$ & $0.059\pm0.021$ & $\ast 18.001\pm6.235$ \\ 
HV 821		& 2.106 & $0.070\pm0.005$ & $4.010\pm0.435$ & $0.071\pm0.009$ & $15.023\pm1.915$ \\
HV 829		& 1.926 & $0.391\pm0.007$ & $4.675\pm0.409$ & $0.031\pm0.042$ & $\ast 15.066\pm42.719$ \\ 
NGC6822-1	& 2.093 & $0.345\pm0.004$ & $4.298\pm0.073$ & $0.086\pm0.004$ & $14.215\pm0.575$  \\
NGC55-1		& 2.245 & $0.185\pm0.024$ & $2.540\pm3.850$ & $0.048\pm0.069$ & $\ast 17.758\pm30.560$ \\ 
NGC55-2		& 2.182 & $0.227\pm0.108$ & $\ast 3.934\pm23.429$ & $0.182\pm0.219$ & $13.983\pm8.321$ \\ 
NGC55-3		& 2.052 & $0.139\pm0.010$ & $4.665\pm0.635$ & $0.062\pm0.017$ & $\ast 16.176\pm4.574$ \\
NGC55-4		& 1.990 & $0.160\pm0.087$ & $\ast 4.380\pm4.331$ & $0.149\pm0.104$ & $15.050\pm5.481$  \\
NGC55-5		& 1.930 & $0.417\pm0.120$ & $4.401\pm1.088$ & $0.053\pm0.094$ & $\ast 14.927\pm34.289$ \\
NGC300-1	& 2.064 & $0.339\pm0.022$ & $4.467\pm0.312$ & $0.093\pm0.014$ & $14.352\pm1.767$ \\
NGC300-2	& 1.950 & $0.310\pm0.055$ & $4.035\pm3.804$ & $0.211\pm0.092$ & $14.703\pm2.784$ \\
NGC300-3	& 1.919 & $0.289\pm0.169$ & $\ast 4.278\pm5.438$ & $\ast 0.329\pm0.607$ & $\ast 13.559\pm9.979$ \\ 
V1		& 2.113 & $0.072\pm0.027$ & $5.354\pm1.773$ & $0.166\pm0.033$ & $18.158\pm1.441$ \\
V15		& 2.097 & $0.102\pm0.032$ & $4.475\pm2.454$ & $0.035\pm0.083$ & $\ast 14.684\pm69.933$ \\ 
\hline
\label{tab:param}
\end{tabular}
\\
Note: $P$ in column 2 represents the period. Entries with $\ast$ are not included in Figure \ref{fig:fp}.
\end{table*}

\begin{figure*}
\centering
\includegraphics[width=0.95\textheight, angle=-90]{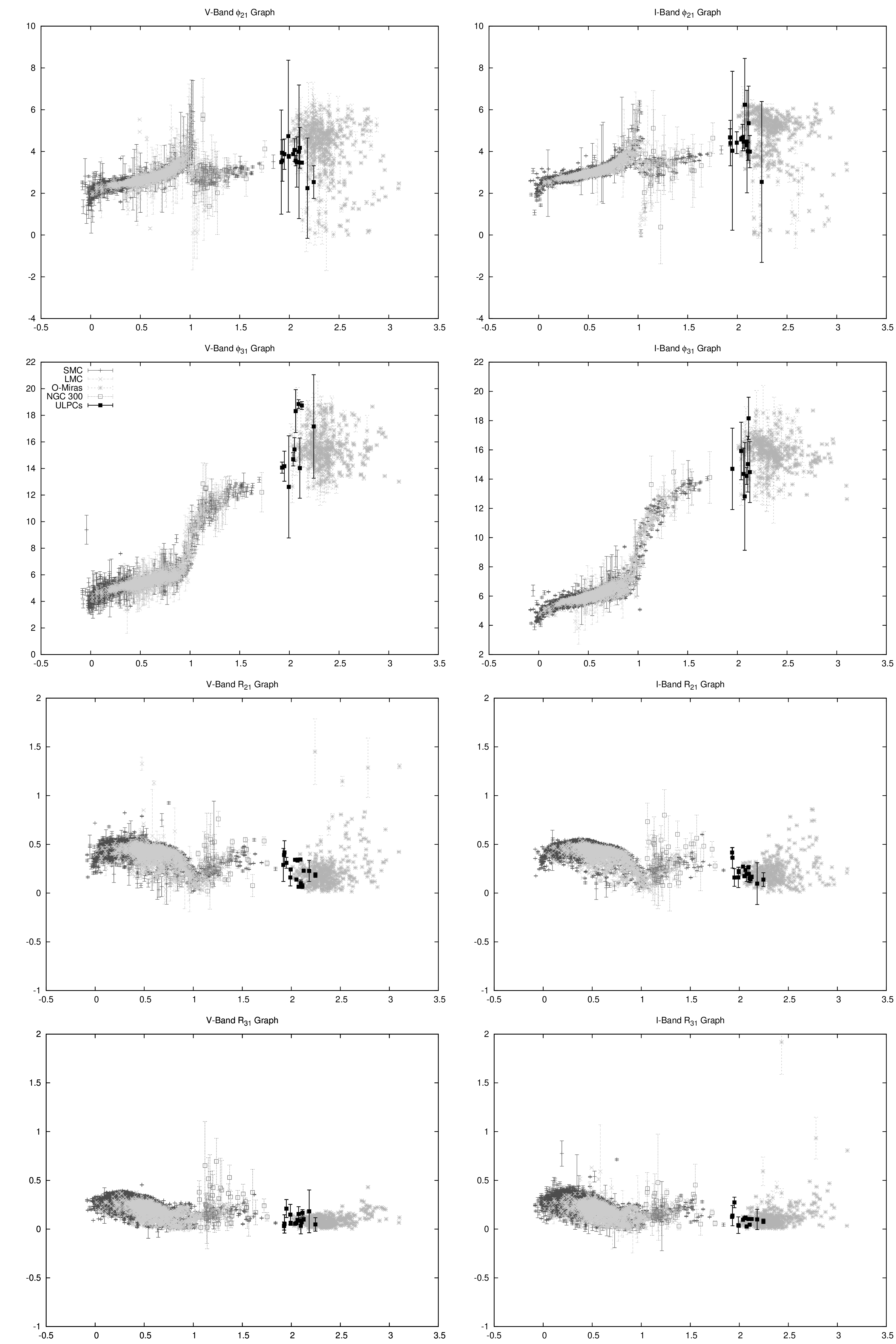}
\caption{Graphs of the Fourier parameters as calculated by the formulas stated in Section \ref{sec:fourier}.  The x-axis is the base 10 logarithm of the period of the star in days, and the y-axis is the value of the various Fourier parameters.  The black points are the ULPCs with the gray points being the Classical Cepheids and Miras from the galaxies stated in Section \ref{sec:data}.  All the data also have error bars as calculated using the method developed by \cite{petersen} described in Section \ref{sec:fouriererr}.}
\label{fig:fp}
\end{figure*}

\subsection{Fourier Parameters}
\label{sec:fourierresults}
Our main goal was to compare the Classical Cepheids, ULPCs and Miras through their light curves structure, by using Fourier analysis technique (see, for example, \cite{simon}).  Figure \ref{fig:fp} shows plots of the $\phi_{21}$, $\phi_{31}$, $R_{21}$, and $R_{31}$ Fourier parameters in both the V- and I-bands, where the values for ULPCs are given in Table \ref{tab:param}. In this Figure, the solid black squares are the ULPCs, and the various shades of gray are the Classical Cepheids and the Miras from their respective galaxies.  Although the error bars are large, and there are many outliers in the Mira data, there does seem to be a trend in the LMC and SMC Classical Cepheids that leads to the ULPCs and beyond to the Miras fairly continuously.
\par
We were able to reconstruct well known features of these Fourier parameter plots (\cite{simon}). In both the V- and I-band graphs of $\phi_{21}$, the resonance at a value of $\log(P)=1$, or at period $P$ of $10$ days, is clearly visible.  After this dramatic drop in the values of $\phi_{21}$, the Classical Cepheids seem to lie along a fairly flat relation, and although we don't have much data between $1.8<\log(P)<2.0$, the ULPCs seem to follow a similar trend. We see similar results for the other Fourier parameter graphs.  For example, in the $\phi_{31}$ graphs we again see a resonance at $\log(P)\sim 1$, then a flattening of the relation (though in this case, the ULPCs do not quite line up as well). Even though there is not much data for stars between $1.8<\log(P)<2.0$, and the error bars on most of the data are quite large, Fourier parameters ($R_{21}$, $R_{31}$, $\phi_{21}$ \& $\phi_{31}$) for ULPCs seem to be extending from the trends defined by Classical Cepheids, and overlapped with the parameters space occupied by Miras.

\section{Discussion \& Conclusion}

\begin{figure*}
\centering
\includegraphics[height=\columnwidth]{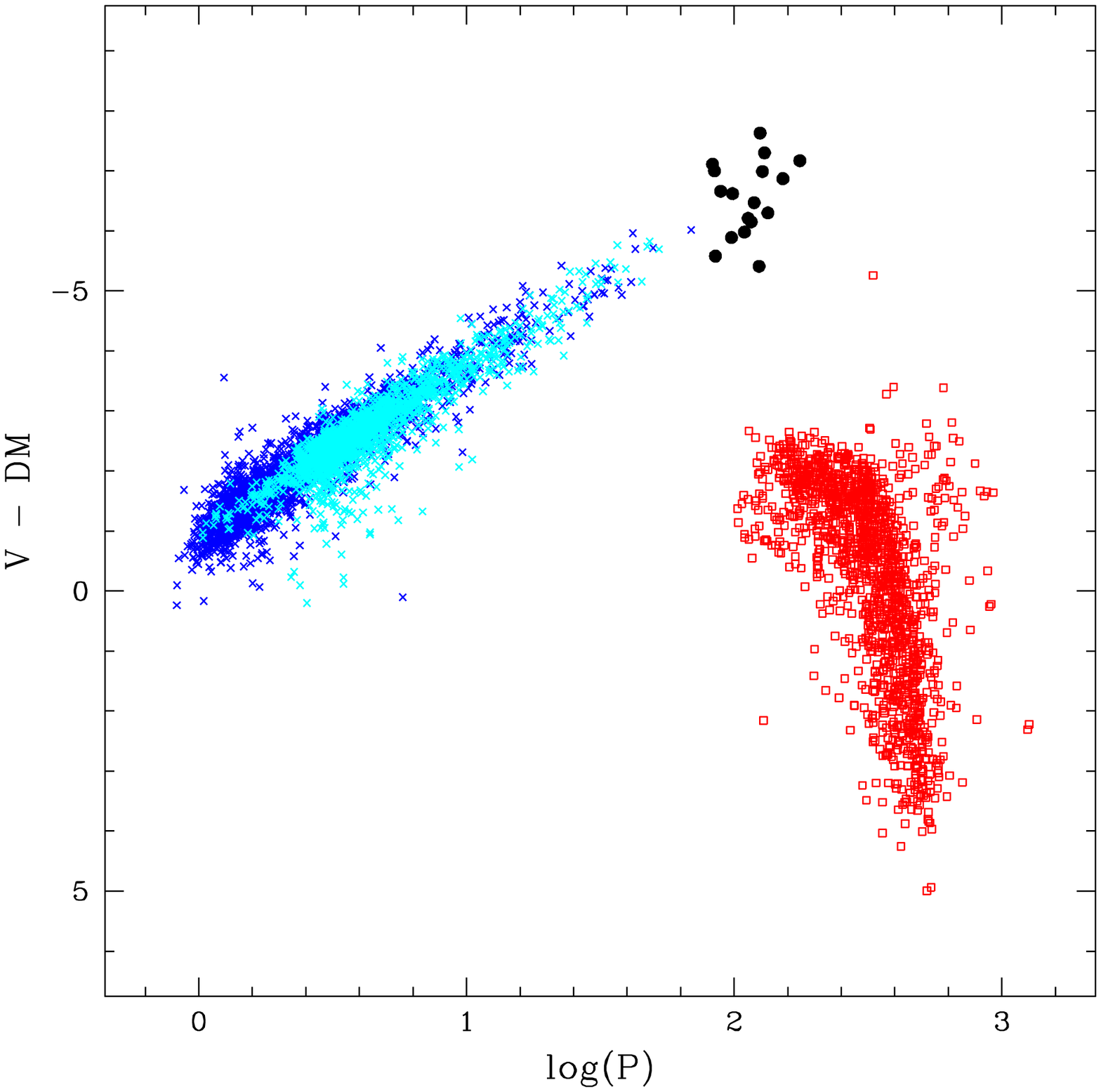}
\includegraphics[height=\columnwidth]{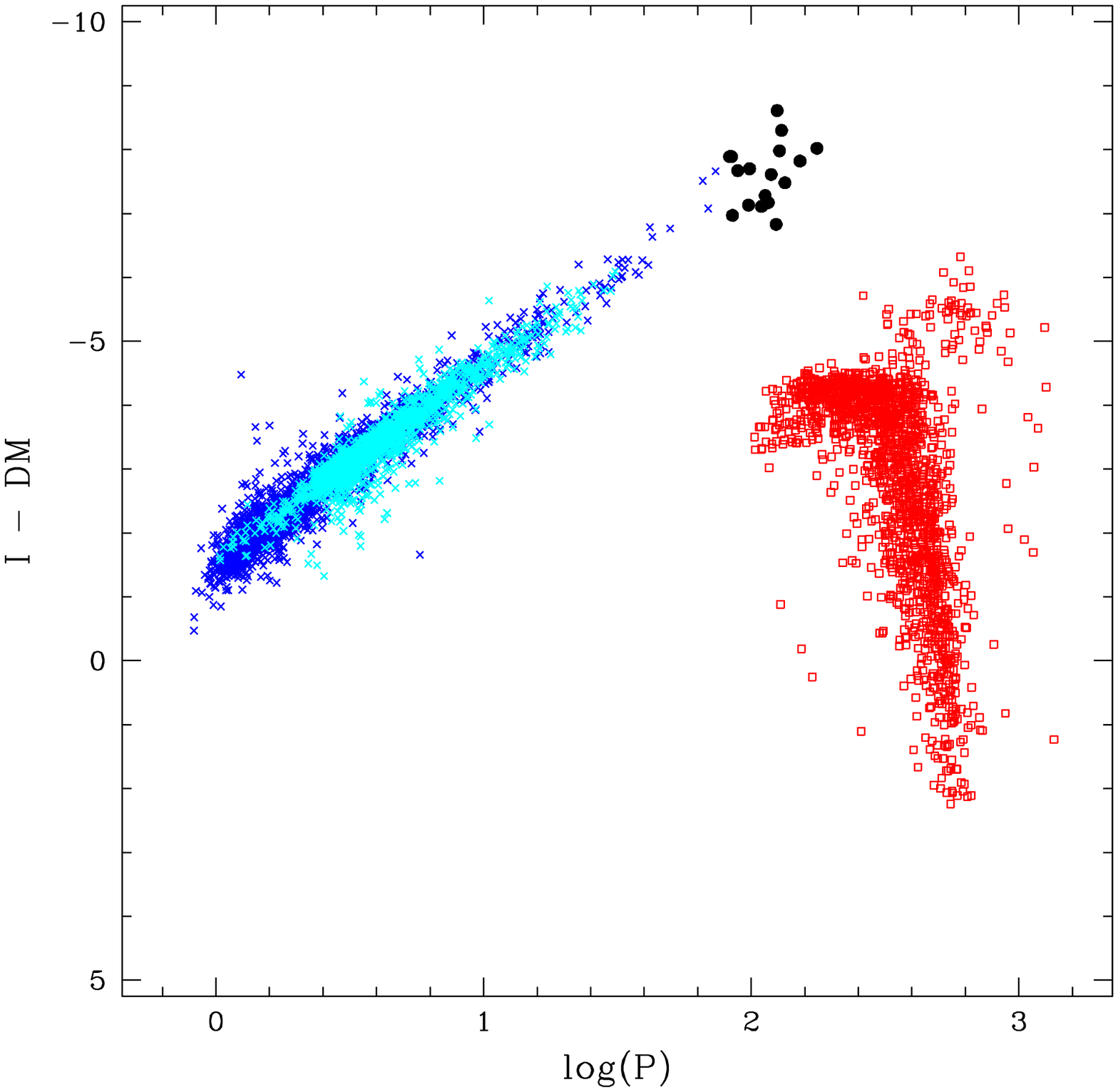}
\caption{Distance modulus (DM; the distance modulus for different galaxies are given in Table \ref{tab:ulpcs}) corrected V- and I-band PL relations. The LMC and SMC Cepheids are shown as crosses in cyan (or light) and blue (or dark) colors, respectively. The Miras data are displayed as (red) open squares. The black filled circles are the data for ULPCs compiled in Section \ref{sec:data}, which are clearly an extension of the PL relations defined by the Classical Cepheids.}
\label{fig:plcompare}
\end{figure*}

By analyzing the light curves of Ultra Long Period Cepheid variable stars, we looked for a relation between shorter period variables (Classical Cepheids) and longer period variables (Miras).  We used Fourier analysis to quantitatively measure the structural properties of the ULPCs' light curves and compare them to the light curves of Classical Cepheids as well as to long period Mira variables in the LMC. It is interesting to take note of the locations of the ULPCs more specifically, with respect to the trends produced by the Miras and the Classical Cepheids. In the graphs of the Fourier parameters (Figure \ref{fig:fp}), the ULPCs seem to be more consistent with the Miras.  They lie more with the longer period variables than along the relations created by the Classical Cepheids. This suggests that when using light curves information to classify variable stars with period longer than 80 days, it is possible that some ULPCs might mis-classified as Miras (and vice versa) due to their overlapped Fourier parameters. In contrast, the plots of period and luminosity can be used to distinguish Classical Cepheids and ULPCs from Miras, as they are well separated in the PL relation (see Figure \ref{fig:plcompare}). Unfortunately, the error bars on Figure \ref{fig:fp} are quite large, and we have a very small sample of ULPCs.   However, once we obtain more data points both for the light curves of known ULPCs and for many more as of yet undiscovered ULPCs, we will be able to learn much more about how it relates to currently known types such as Classical Cepheids and Mira variables.

\section*{Acknowledgment}
The authors thank SUNY Oswego, National Central University (NCU) and the Graduate Institute for Astronomy at NCU, and the National Science Foundation's Office of International Science \& Engineering's award 1065093.  CCN thanks the funding from National Science Council of Taiwan under the contract NSC101-2112-M-008-017-MY3.  We would also like to thank G. Fiorentino for her helpful comments when drafting this paper.  Finally, this project wouldn't have been possible without the ULPC data compiled by J. Bird and G. Fiorentino.


\begin{thebibliography}{1}

\bibitem{bird} Bird, J.~C., Stanek, K.~Z., \& Prieto, J.~L.\ 2009, The Astrophysical Journal, 695, 874 

\bibitem{moffett} Moffett T., Gieren W. P., Barnes T. G. III \& Gomez M., 1998, The Astrophysical Journal, 117, 135

\bibitem{freedman} Freedman W. L., Grieve G. R. \& Madore B. F., 1985, The Astrophysical Journal, 59, 311

\bibitem{martin} Martin W. L. \& Warren P. R., 1979, South African Astronomical Observatory Circulars, 4, 98

\bibitem{fiorentino} Fiorentino G. et al., 2010, The Astrophysical Journal, 711, 808

\bibitem{soslmc} Soszy\'nski I. et al., 2008, Acta Astronomica, 58, 163

\bibitem{sossmc} Soszy\'nski I. et al., 2010, Acta Astronomica, 60, 17

\bibitem{sos} Soszy\'nski I. et al., 2009, Acta Astronomica, 59, 239

\bibitem{gieren} Gieren W. et al., 2004, The Astrophysical Journal, 128, 1167

\bibitem{petersen} Petersen J. O., 1986, Astronomy and Astrophysics, 170, 59

\bibitem{simon} Simon N. R. \& Lee A. S., 1981, The Astrophysical Journal, 248, 291

\end{thebibliography}
\end{document}